\input{aipcheck}

\documentclass[final,sort&compress]{aipproc}
\usepackage{amsfonts}
\usepackage{amsmath}
\usepackage{amssymb}
\usepackage{graphicx}

\layoutstyle{6x9}

\begin{document}

\title{Nonextensive statistical mechanics and central limit theorems II - 
Convolution of $q$-independent random variables}


\author{S\'{\i}lvio M. Duarte~Queir\'{o}s}{ 
  address={Centro Brasileiro de Pesquisas F\'{\i}sicas, 150, 22290-180, Rio de Janeiro
- RJ, Brazil}
}

\author{Constantino~Tsallis}{
  address={Centro Brasileiro de Pesquisas F\'{\i}sicas, 150, 22290-180, Rio de Janeiro                                        
- RJ, Brazil}, altaddress={Santa Fe Institute, 1399 Hyde Park Road, Santa Fe, New Mexico 87501, USA}                                                                 
}


\begin{abstract}

In this article we review recent generalisations of the central limit
theorem for the sum of specially correlated (or \textit{$q$-independent})
variables, focusing on $q \ge 1$. Specifically, this kind of correlation turns the probability
density function $\mathcal{G}_{q}\left( X\right) =\mathcal{A}_{q}\left[
1+\left( q-1\right) \beta _{q}\left( X-\bar{\mu}_{q}\right) ^{2}\right]
^{\frac{1}{1-q}}$, which emerges upon maximisation of the entropy $%
S_{q}=k\left( 1-\int \left[ p\left( X\right) \right] ^{q}\ dX\right) /\left(
1-q\right) $, into an attractor in probability space. Moreover, we also discuss
a $q$-generalisation of $\alpha $-stable L\'{e}vy distributions which can
as well be stable for this special kind of correlation. Within this context, we
verify the emergence of a triplet of entropic indices which relate
the form of the attractor, the correlation, and the scaling rate,
similar to the $q$-triplet that connects the entropic indices characterising
the sensitivity to initial conditions, the stationary state, and relaxation to
the stationary state in anomalous systems.

\end{abstract}

\maketitle

\section{Introduction}

\label{intro}

In our previous article \cite{ct-smdq-1}, we have verified the standard central limit theorem and its L\'{e}vy-Gnedenko extension
for the case of independent and identically distributed random variables
associated with a $q$-Gaussian distribution,%
\begin{equation}
\mathcal{G}_{q}\left( x\right) \equiv \mathcal{A}_{q}\,e_{q}^{-\beta %
_{q}\left( x-\bar{\mu}_{q}\right) ^{2}},  \label{pq}
\end{equation}%
with $e_{q}^{x}\equiv \left[ 1+\left( 1-q\right) \,x\right] ^{\frac{1}{1-q}}$ (if $1+(1-q)x \ge 0$, and zero otherwise)
$\left( e_{1}^{x}\equiv e^{x}\right) $. Distribution (1) optimises the continuous version of the nonadditive entropy 
\cite{ct88} \ $S_{q}\equiv k\left( 1-\sum\limits_{i=1}^{W}p_{i}^{q}\right)
/\left( 1-q\right) $, where $q\in \Re $. In this article we review recent
generalisations of the central limit theorem which have been formulated
within nonextensive statistical mechanical concepts. After numerical
indications suggesting the existence of a generalisation, within nonextensive
statistical mechanics, of the central limit theorem~\cite{goyo-ct-gm,thi}
~\footnote{
The models introduced in references~\cite{goyo-ct-gm} and~\cite{thi} have recently been analytically
solved in Ref.~\cite{hilhorst}. It was verified that, for these two
cases, the limiting distributions are {\it not} in fact  $q$-Gaussians, but are instead other distributions. 
These distributions are, however, so close to $q$-Gaussians \textquotedblleft that
it becomes extremely difficult to distinguish the true curve from its
q-Gaussian approximant\textquotedblright ~\cite{hilhorst}. These interesting results show that the probabilistic correlations included in these two specific models do  {\it not} correspond exactly to {\it $q$-independence}, but are only very close to it instead.}, such
generalisation has indeed been proved for variables with finite or
 infinite $q$-generalized second-order moment \cite{q-clt-gauss,q-clt-levy}.

\section{Central limit theorems for $q$-independent variables}

\subsection{The $q$-Gaussian case}

As we have reviewed and illustrated, in the case of convolution of
{\it independent} random variables (including random variables associated with $%
\mathcal{G}_{q}\left( X\right) $ distributions), there are only two stable
forms in probability space, namely, the Gaussian and the $\alpha $%
-stable L\'{e}vy probability density functions. We have also referred in Part
I \cite{ct-smdq-1} to other versions of the central limit theorem that are available in the literature. In what follows we discuss the
addition of random variables that are correlated in such a special way that
a new algebra, the $q$-algebra \cite{nivanen,borges}, is necessary for
a suitable analysis. In this context, it has been introduced in Ref.~\cite%
{q-clt-gauss} a non-linear integral transform, the $q$-Fourier Transform, 
\begin{equation}
\mathcal{F}_{q}\left[ f\right] \left( k\right) \equiv \int\nolimits_{-\infty
}^{\infty }e_{q}^{i\,k\,X}\otimes _{q}f\left( X\right) \,dX.
\label{q-fourier}
\end{equation}

Applying the definition of $q$-product, $x\otimes _{q}y\equiv \left[
x^{1-q}+y^{1-q}-1\right] ^{\frac{1}{1-q}}$, in Eq.~(\ref{q-fourier}), it is
possible to write $\mathcal{F}_{q}\left[ f\right] \left( k\right) $ without
the explicit use of the $q$-product, $\mathcal{F}_{q}\left[ f\right] \left(
k\right) =\int\nolimits_{-\infty }^{\infty }f\left( X\right) \exp _{q}\left[ 
\frac{ikX}{\left\{ f\left( X\right) \right\} ^{1-q}}\right] \,dX$. As an
application of $\mathcal{F}_{q}\left[ f\right] \left( k\right) $, it is
provable that,%
\begin{equation}
\mathcal{F}_{q}\left[ \mathcal{G}_{q}\left( X\right) \right] \left( k\right)
=\left\{ \exp _{q}\left[ -\frac{k^{2}}{4\beta ^{2}-q\,C_{q}^{2\left(
q-1\right) }}\right] \right\} ^{\frac{3-q}{2}},  \label{qgauss-q-fourier}
\end{equation}
where $C_{q}=\sqrt{\beta }/\mathcal{A}_{q}$, and $\beta = \mathcal{B}$  with $\mathcal{A} $ and $\mathcal{B} $ 
as defined in Eq.~(9) of Part~I~\cite{ct-smdq-1}.

Another distribution for which it is simple to obtain its $q$-Fourier
Transform, is the uniform distribution, $\mathcal{U}\left( X\right) $, $%
\mathcal{U}\left( X\right) =\frac{1}{2a}$ ($-a\leq X\leq a$, $a>0$) 
\footnote{%
In this case, as well as for all distributions with compact support, integration must be done over
that support. Otherwise the integral does not converge.}. Its $q$-Fourier
Transform is, $\mathcal{F}_{q}\left[ \mathcal{U}\left( X\right) \right]
\left( k\right) =\frac{\check{q}^{2}\left( 2\,a\right) ^{\left( \check{q}%
-1\right) /\check{q}}}{a\,k}\sin _{\check{q}}\left[ \frac{a\,k}{\check{q}%
\left( 2\,a\right) ^{\left( \check{q}-1\right) /\check{q}}}\right] $, where $%
\sin _{q}\left( x\right) $ represents the $q$-generalisation of $\mathrm{%
\sin }\left( x\right) $~\cite{q-seno}, and $q=2-\frac{1}{\check{q}}$. 

\begin{figure}[tbh]
\includegraphics[width=0.5\columnwidth,angle=0]{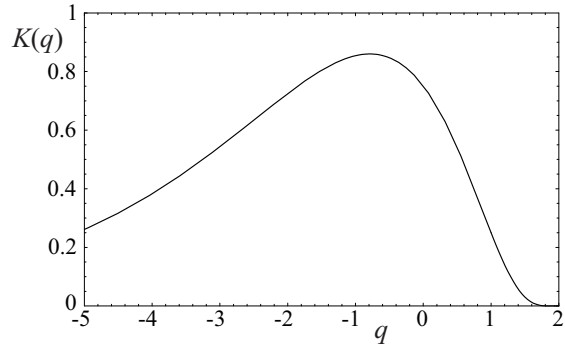} 
\caption{Representation of Eq.~(\protect\ref{q-incerteza}) for values of $q$
between $-5$ and $2$. For $q=1$ $K(q)=1/4$; $\lim_{q \to -\infty} K(q)=0$. We are presently focusing on $q \ge 1$. 
}
\label{fig-incerteza}
\end{figure}

Introducing a
function,%
\begin{equation}
v\left( u\right) =\frac{1+u}{3-u},\qquad (u<3),  \label{map}
\end{equation}%
whose inverse is,%
\begin{equation}
v^{-1}\left( u\right) =\frac{3u-1}{1+u},\qquad (u>-1),  \label{invmap}
\end{equation}%
and assuming $q_{1}=v\left( q\right) $ and $q_{-1}=v^{-1}\left( q\right) $,
it is possible to rewrite  Eq.~(\ref{qgauss-q-fourier}) as follows:%
\begin{equation}
\mathcal{F}_{q}\left[ \mathcal{G}_{q}\left( X\right) \right] \left( k\right)
=\exp _{q_{1}}\left[ -\beta _{q}^{\prime }\,k^{2}\right] ,
\label{q-fourier-1}
\end{equation}%
and $\mathcal{F}_{q_{-1}}\left[ \mathcal{G}_{q_{-1}}\left( X\right) \right]
\left( k\right) =\exp _{q}\left[ -\beta _{q_{-1}}^{\prime }\,k^{2}\right] ,$%
where 
\begin{equation}
\beta _{q}^{^{\prime }}=\frac{3-q}{8\,\beta ^{2-q}\,C_{q}^{2\left(
q-1\right) }}.  \label{beta-fourier}
\end{equation}%
Equation (\ref{beta-fourier}) can be rewritten as%
\begin{equation}
\left[ \beta _{q}^{^{\prime }}\right] ^{\frac{1}{\sqrt{2-q}}}\,\beta ^{\sqrt{%
2-q}}=\left( \frac{3-q}{8\,C_{q}^{2\left( q-1\right) }}\right) ^{\frac{1}{%
\sqrt{2-q}}}\equiv K\left( q\right) ,\qquad q<2.  \label{q-incerteza}
\end{equation}%
We might consider the case $q=1$ in Eq. (\ref{q-incerteza}). In this
situation, the Fourier Transform of $\mathcal{G}_{1}\left( X\right) $ has
the same functional form, $\mathcal{G}_{1}\left( k\right) =\exp \left[
-\beta ^{\prime }\,k^{2}\right] $. For Gaussian functions like these
two, the width (and the inflexion point in linear-linear scale) are related
to $\beta $ (actually $\frac{1}{\sqrt{2\,\beta }}$). Hence, Eq. (\ref%
{q-incerteza}) reflects a relation between uncertainties in real and
reciprocal spaces. In the general case,  relation (8) is a sort of $q$-analogue of the
quantum mechanical uncertainty principle by \textsc{Werner Heisenberg} \cite%
{heisenberg}. In Fig.~\ref{fig-incerteza} we represent  $K(q)$
for values of $q$ between $-5$ and $2$.

Equation (\ref{q-fourier-1}) has permitted to
verify the mapping, through $\mathcal{F}_{q}$,%
\begin{align}
\mathcal{G}_{q}\underset{\mathcal{F}_{q}}{\rightarrow }\mathcal{G}%
_{q_{1}},\qquad q_{1}& =v\left( q\right) ,\quad 1 \le q  < 3,
\label{q-fourier-transf} \\
\mathcal{G}_{q_{-1}}\underset{\mathcal{F}_{q_{-1}}}{\rightarrow }\mathcal{G}%
_{q},\qquad q_{-1}& =v^{-1}\left( q\right) ,\quad 1 < q,  \notag
\end{align}%
and to prove the existence of the \emph{inverse }$q$\emph{-Fourier Transform}%
, $\mathcal{F}_{q}^{-1}$, $\mathcal{G}_{q_{1}}\underset{\mathcal{F}_{q}^{-1}}%
{\rightarrow }\mathcal{G}_{q}$ ($q_{1}=v\left( q\right) ,\quad 1<q<3$), and $%
\mathcal{G}_{q}\underset{\mathcal{F}_{q_{-1}}^{-1}}{\rightarrow }\mathcal{G}%
_{q_{-1}}$ ($q_{-1}=v^{-1}\left( q\right) ,\quad 1<q$). It is worth to
mention that $q_{1}$ and $q_{-1}$ fulfil the dual relation $q_{-1}+\frac{1}{%
q_{1}}=2$, that has also appeared in the context of phase space
self-invariant occupancy~\cite{ct-gm-sato}, {\it i.e.}, such as that all marginal probabilities of the system composed by 
$N$ equal and distinguishable subsystems coincide or asymptotically approach (for $N \rightarrow \infty $) 
the joint probabilities of the $(N - 1)$-system. If we consider a sequence of
applications of $v\left( q\right) $, $q_{n}=v_{n}\left( q\right) $, it can be seen that 
\begin{equation}
\frac{2}{1-q_{n}}=\frac{2}{1-q}+n,\qquad n=0,\pm
1,\pm 2,\ldots \mathrm{\ and}\quad q=v_{0}\left( q_{0}\right) .
\label{qn-sucessao}
\end{equation}%
If $q=1$, then $q_{n}=1$ for all $n$. Otherwise, i.e. if $q\neq 1$, in the limit $%
n\rightarrow \pm \infty $, $q_{n} \to 1$. This result can be interpreted in the
following way: \emph{the simple successive application of the }$q$\emph{%
-Fourier\ Transform,} 
\begin{equation}
\mathcal{F}_{q_{n}}^{m}\equiv \mathcal{F}_{q_{n+m-1}}\circ \ldots \circ 
\mathcal{F}_{q_{n}},\qquad m=1,2,\ldots \mathrm{\ and\quad }n=0,\pm 1,\pm
2,\ldots 
\end{equation}
on a distribution $\mathcal{G}_{q}$ leads to the Gaussian form, $\underset{%
m\rightarrow \pm \infty }{\lim }\mathcal{F}_{q}^{m}\left[ \mathcal{G}_{q}%
\right] =\mathcal{G}$.

Let us now present the scheme within which the functional form Eq.~(\ref{pq}), 
\textit{i.e.}, $\mathcal{G}_{q}\left( X\right) $\emph{, turns out to be
stable}. 

\emph{Two random variables, }$X$\emph{\ and }$Z$\emph{,
are said }$q$\emph{-independent if} 
\begin{equation}
\mathcal{F}_{q_{\ast }}\left[ \mathcal{P}\left( X+Z\right) \right] \left(
k\right) =\mathcal{F}_{q_{\ast }}\left[ p\left( X\right) \right] \left(
k\right) \otimes _{q}\mathcal{F}_{q_{\ast }}\left[ p^{\prime }\left(
Z\right) \right] \left( k\right) ,  \label{q-convolution}
\end{equation}%
where $q_{\ast }=q_{-1}$, $\mathcal{P}\left( X+Z\right) $, $p\left( X\right) 
$, and $p^{\prime }\left( Z\right) $ are the probability density functions
for $X+Z$, $X$ and $Z$, respectively. As an exemple of $q$-independency we
refer two variables, $X$ and $Z$, both associated with a $q_{\ast }$%
-Gaussian probability density function, $\mathcal{G}_{q_{\ast }}\left(
X\right) $ and $\mathcal{G}_{q_{\ast }}\left( Z\right) $, with $\beta _{X}$
and $\beta _{Z}$, respectively. If the variables are $q$-independent then
the sum $X+Z$ is also associated with a $q_{\ast }$-Gaussian whose width is $%
\delta =\left( \frac{3-q}{8\left( \gamma _{X}+\gamma _{Y}\right)
C_{q}^{2\left( q-1\right) }}\right) ^{\frac{1}{2-q}}$, where $\gamma
_{\left( .\right) }=\frac{3-q}{8\,\beta _{\left( .\right)
}^{2-q}\,C_{q}^{2\left( q-1\right) }}$.

Consider now the variable, $Y\equiv \frac{X_{1}+X_{2}+\ldots +X_{N}-N\,\mu
_{q_{\ast }}}{D_{N}\left( q_{\ast }\right) }$, where $X_{i}$, $i\in \left(
1,N\right) $, are identically distributed random variables.

Let $X_{1},\ldots ,X_{N}$ be a sequence of $q$-independent identically
distributed random variables with finite $q_{\ast }$-mean, $\mu _{q_{\ast }}$%
, and finite second $\left( 2q_{\ast }-1\right) $-order moment, $\sigma
_{2q_{\ast }-1}^{2}$. Under these conditions, with 
\begin{equation}
D_{N}\left( q_{\ast }\right) \equiv \left( \sqrt{N\,\mathcal{Z}_{2q-1}}%
\sigma _{2q_{\ast }-1}\right) ^{\frac{1}{2-q_{\ast }}},  \label{x-scaling}
\end{equation}%
$\mathcal{P}\left( Y\right) $ is said to be $q$-convergent to a $q_{-1}$%
-normal distribution as $N\rightarrow \infty $, with $\mathcal{Z}%
_{u}=\int\nolimits_{-\infty }^{\infty }\left[ f\left( X\right) \right]
^{u}\,dX$. Analytically, this can be written as,%
\begin{equation}
\mathcal{F}_{q_{\ast }}\left[ \mathcal{P}\left( Y\right) \right] \left(
k\right) =\mathcal{G}_{q}\left( k\right) =\mathcal{F}_{q_{-1}}\left[ 
\mathcal{G}_{q_{-1}}\left( X\right) \right] \left( k\right) .  \label{clt-ct}
\end{equation}%
The proof of Eq. (\ref{clt-ct}) \cite{q-clt-gauss} follows along the lines
of the standard central limit theorem where, instead of $\mathcal{F}%
\left[ f\right] \left( k\right) $ and the usual product, we use $\mathcal{F}%
_{q_{\ast }}\left[ f\right] \left( k\right) $ with
\begin{equation}
\mathcal{F}_{q_{\ast }}\left[ f\right] \left( k\right) =1+i\mu _{q_{\ast }}\,%
\mathcal{Z}_{q_{\ast }}\,k-\frac{q_{\ast }}{2}\,\,\sigma _{2q_{\ast
}-1}^{2}\,\mathcal{Z}_{2q_{\ast }-1}\,k^{2}+\mathcal{O}\left( k^{2}\right)
,\qquad \left( k\rightarrow 0\right) ,  \label{clt1}
\end{equation}%
and the $q$-product. Thus, the convolution of $q$-independent random variables $Y$ is written as,%
\begin{equation}
\mathcal{F}_{q_{\ast }}\left[ \mathcal{P}\left( Y\right) \right] \left(
k\right) =\mathcal{F}_{q_{\ast }}\left[ p\left( X\right) \right] \left(
k\right) \otimes _{q}\ldots \otimes _{q}\mathcal{F}_{q_{\ast }}\left[
p\left( X\right) \right] \left( k\right) ,\qquad \mathrm{(N\ factors)},
\label{clt2}
\end{equation}%
where $1 \le q\leq 2$ and consequently $1\le q_{\ast }\leq \frac{5}{3}$.

Using Eq. (\ref{clt2}) in Eq. (\ref{clt1}), together with properties of $q$%
-logarithm when $k\rightarrow 0$ , $\ln _{q}\,\mathcal{F}_{q_{-1}}\left[ 
\mathcal{P}\left( Y\right) \right] \left( k\right) =N\,\ln _{q}\left( 1-%
\frac{q_{-1}}{2}\,\,\frac{\,k^{2}}{N}+\mathcal{O}\left( \frac{k^{2}}{N}%
\right) \right) $, \textit{i.e.}, 
\begin{equation}
\underset{N\rightarrow \infty }{\lim }\mathcal{F}_{q_{-1}}\left[ \mathcal{P}%
\left( Y\right) \right] \left( k\right) =\exp _{q}\left[ -\frac{q_{-1}}{2}%
k^{2}\right] .  \label{clt-ct-final}
\end{equation}%
In other words, $Y$ \emph{\ is }$q_{-1}$\emph{-convergent to the random
variable }$Z$\emph{\ whose }$q_{-1}$\emph{-Fourier Transform is given by Eq.~%
}(\ref{clt-ct-final}). Hence, according to the mapping relations, the explicit
form of the corresponding $q_{-1}$-Gaussian, $\mathcal{G}_{q_{-1}}\left(
X\right) $, yields%
\begin{equation}
\mathcal{G}_{q_{-1}}\left( X\right) =\frac{1}{\mathcal{Z}_{q_{-1}}}\frac{%
\beta _{q_{-1}}}{\left( 2\sqrt{1+q-1}\right) ^{1/\left( 2-q_{-1}\right) }}%
\exp _{q_{-1}}\left[ -\beta _{q_{-1}}\,X^{2}\right]  \,,
\end{equation}%
where $\beta _{q_{-1}}\equiv \left[ \left( 3-q_{-1}\right) /\left( 4\,q\,%
\mathcal{Z}_{q_{-1}}^{2\left( q_{-1}-1\right) }\right) \right] ^{^{\frac{1}{%
2-q_{-1}}}}$. Obviously, when we $q$-convolute $q_{\ast }$-Gaussians, the
resulting probability density function is a $q_{\ast }$-Gaussian, \textit{%
i.e.}, $\mathcal{G}_{q_{\ast }}\left( X\right) $ \emph{is a stable attractor
for }$q$\emph{-independent random variables upon the condition stated above},
in the same way $\mathcal{G}\left( X\right) $ and $L_{\alpha }\left( X\right) $ 
are the stable attractors for the sum of {\it independent} random variables.

Concerning the exponent of Eq.~(\ref{x-scaling}), it is easy to verify that $%
\left( 2-q_{\ast }\right) ^{-1}=v\left( v\left( q_{-1}\right) \right) \equiv
q_{1}$. Defining $\delta \equiv \left( 2-q_{-1}\right) ^{-1}$, as the \emph{%
scaling rate exponent}, we have $\delta =q_{1}$. We are then allowed to
define, for generic $n$ (following Eq.~(\ref{qn-sucessao})), a $q$-triplet, $%
\left\{ q_{att},q_{corr},q_{scal}\right\} $, which relates entropic indices for the
attractor, $q_{att}=q_{n-1}$, the correlation, $q_{corr}=q_{n}$, and the scaling, 
$q_{scal}=q_{n+1}$.

Consider now the variable $Y^{\prime }=Y\,D_{N}\left( q_{\ast }\right) $,
that is composed by the sum of $N$ $q$-independent random variables all
following a $q_{\ast }$-Gaussian with the same $\beta _{q_{\ast }}$. Using
the associative property of the $q$-product in Eq. (\ref{clt2}), we are
able to obtain the scaling relation,%
\begin{equation}
\beta _{q_{\ast }}^{\prime }\left( Y\right) =N\,\beta _{q_{\ast }}^{\prime },
\label{scaling-fourier}
\end{equation}%
in Fourier space (see Eq. (\ref{beta-fourier})). For the standard ($q_{\ast
}=1$) and generalised ($2q_{\ast }-1$)-variances, we have
obtained the relations%
\begin{equation}
\sigma ^{2}\left( Y\right) =N^{\frac{1}{2-q_{\ast }}}\sigma ^{2},\qquad 
\mathrm{and\quad }\sigma _{2\,q_{\ast }-1}^{2}\left( Y\right) =N^{\frac{1}{%
2-q_{\ast }}}\sigma _{2\,q_{\ast }-1}^{2},  \label{scaling-sigma}
\end{equation}%
with $\sigma ^{2}=\left[ \beta _{q_{\ast }}\left( 5-3\,q_{\ast }\right) %
\right] ^{-1}$ and $\sigma _{2\,q_{\ast }-1}^{2}=\left[ \beta _{q_{\ast
}}\left( 1+q_{\ast }\right) \right] ^{-1}$. When $q=1$, Eq. (\ref%
{scaling-sigma}) turns out into the well known relation for the sum of
independent variables, $\sigma ^{2}\left( Y\right) =N\,\sigma ^{2}$. The way on which variance scales upon addition 
is a standard tool to evaluate the character of a time series whose elements $X_{i}$ have a variance $\sigma ^{2}$.
It is well known that the nature of a signal is characterised by its \textit{Hurst exponent}, $H$, 
$\sigma ^{2}_{N} = N^{2H} \sigma ^{2}$ where $\sigma ^{2}_{N}$ represents the variance of a new variable obtained from the sum of
$N$ elements of the time series with variance $\sigma ^{2}$. The series is considered as {\it anti-persistent} if $0<H<1/2$, 
{\it Brownian} if $H=1/2$, and {\it persistent} if $1/2<H<1$ \cite{feder}. By comparing the Hurst exponent definition 
with Eq.~(\ref{scaling-sigma}) we verify that a connection can be established. Expressly, and from Eq. (\ref{scaling-sigma}), 
we conjecture that $q$-independent stochastic signals should respect the following relation 
$H=\left[ 2\left( 2-q_{\ast }\right) \right] ^{-1}$.

\subsubsection{Verification of the $q$-generalised CLT}

We verify here that correlations of the
form of Eq.~(\ref{q-convolution}) yield a $q$-Gaussian attractor.

\begin{figure}[tbh]
\includegraphics[width=0.8\columnwidth,angle=0]{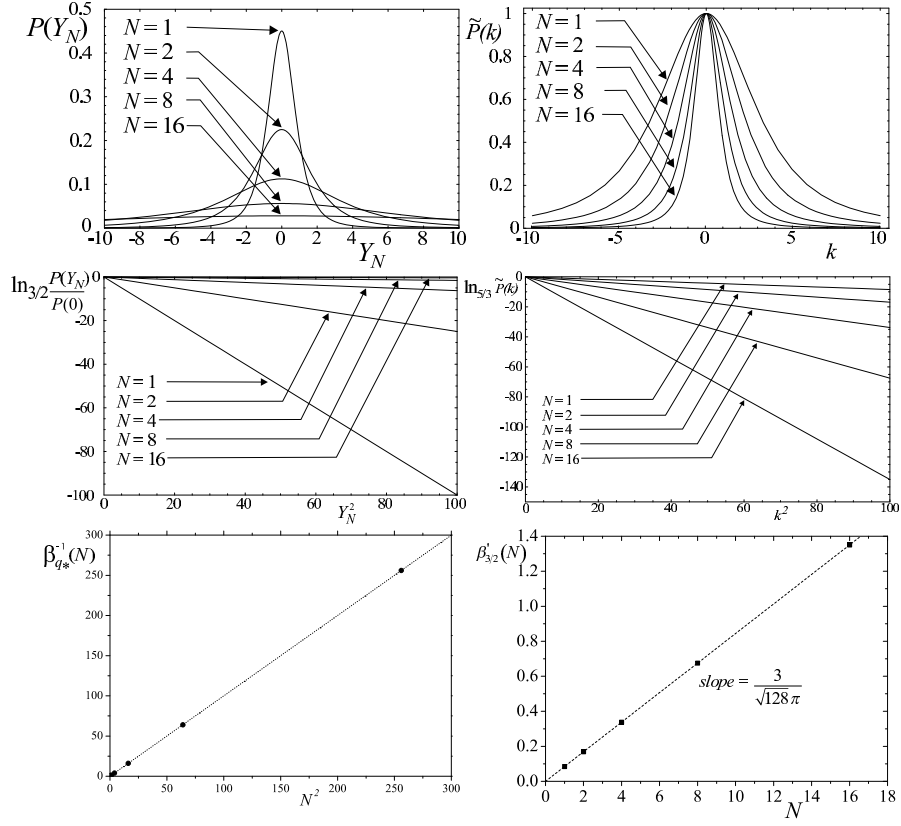} %
\caption{ {\it Upper panels:} Probability distribution $P\left(
Y_{N}\right) $ \textit{vs.} $Y_{N}$, with $Y_{N}\equiv \sum_{i=1}^{N}X_{i}$ ,
$X_{i}$ being $\left( q=\frac{5}{3}\right) $-independent random 
variables associated with a $\mathcal{G}_{\frac{3}{2}}\left( X\right) $
distribution with $\protect\beta =1$ ({\it left}), and the respective $\left( q=%
\frac{3}{2}\right) $-Fourier Transform, $\tilde{P}\left( k\right) $, \textit{%
vs.} $k$ ({\it right}). {\it Middle panels:} Same as above, in $\ln _{\frac{3}{2}}$%
-squared scale ({\it left}), and $\ln _{\frac{5}{3}}$-squared scale ({\it right}).
The straight lines indicate that $P\left( Y_{N}\right) $ and $%
\tilde{P}\left( k\right) $ are $q$-Gaussians with $q=\frac{3}{2}$ and $q=\frac{5}{3}$ respectively.
Their slopes are $\protect\beta _{q_{\ast }=3/2}^{-1}\left( N\right) $ for
left panel curves and $\protect\beta _{q_{\ast }}^{\prime }\left( N\right) $
for right panel curves. {\it Lower panels:} $\protect\beta _{q_{\ast
}=3/2}^{-1}\left( N\right) $ \textit{vs.} $N^{2}$, which is a straight line
with slope $1$ ({\it left}); $\protect\beta _{q_{\ast }=3/2}^{\prime }\left(
N\right) $ \textit{vs.} $N$ which is also a straight line but with slope $%
\left. \frac{3-q_{\ast }}{8\,C_{q_{\ast }}^{2\left( q_{\ast }-1\right) }}%
\right\vert _{q_{\ast }=3/2}=0.088844\ldots $ ({\it right}). }
\label{clt-32}
\end{figure}

Let us start with the case of $q_{\ast }$-Gaussians with $q_{\ast }=3/2$ and 
$\beta =1$ which are $q$-independent. By direct evaluation of its $q_{\ast }$%
-Fourier Transform, $\mathcal{F}_{q_{\ast }}\left[ \mathcal{G}_{q_{\ast
}}\left( X\right) \right] \left( k\right) $, and taking into account the
Cauchy principal value, we obtain 
\begin{equation}
\mathcal{F}_{\frac{3}{2}}\left[ \mathcal{G}_{\frac{3}{2}}\left( X\right) %
\right] \left( k\right) =\left[ 1+\frac{1}{2\sqrt{2}\pi }k^{2}\right] ^{-%
\frac{3}{2}},  \label{qfourier32}
\end{equation}%
which corresponds exactly to a $\mathcal{G}_{\frac{5}{3}}\left( k\right) $
function with the same $q$ and $\beta $ as indicated by Eq.~(\ref{beta-fourier}). Using Eq. (\ref%
{qfourier32}) in Eq. (\ref{q-convolution}), and the mapping relations (\ref%
{map}) and (\ref{invmap}), we have obtained the convolution of two $\mathcal{%
G}_{\frac{3}{2}}\left( X\right) $ distributions which is also a $\frac{3}{2}$%
-Gaussian. From the latter, and applying the associative property of the $q$%
-product, we have calculated the convolution of $N=2,4,8,16$ $\left( q=\frac{%
3}{2}\right) $-Gaussian distributions. Contrarily to what happens for $\left( q=1\right) $%
-independent variables, the resulting distribution is \emph{always} a $\left( q=\frac{%
3}{2}\right) $-Gaussian that will \emph{never} converge to a Gaussian, even
when $N\rightarrow \infty $, and despite the finiteness of the $\mathcal{G}_{%
\frac{3}{2}}\left( X\right) $ standard deviation. On the panels of Fig.~\ref{clt-32} we depict the behaviour we have just described.

\begin{figure}[tbh]
\includegraphics[width=0.8\columnwidth,angle=0]{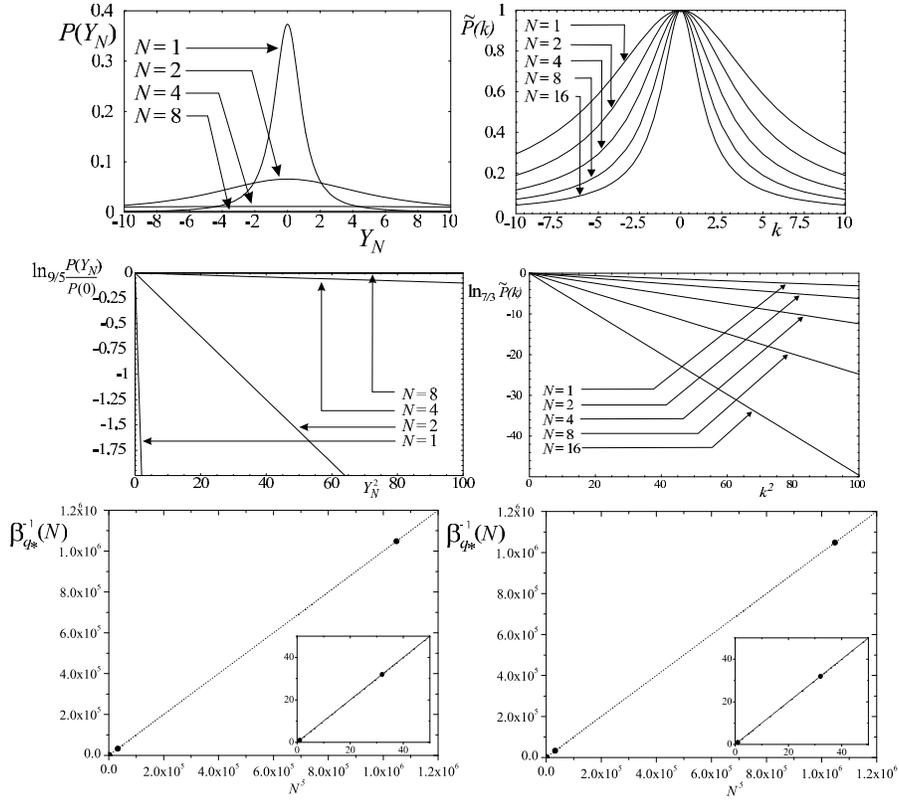} %
\caption{{\it Upper panels:} Probability distributions $P\left(
Y_{N}\right) $ \textit{vs.} $Y_{N}$, with $Y_{N}\equiv \sum_{i=1}^{N}X_{i}$ ,
$X_{i}$ being $\left( q=\frac{7}{3}\right) $-independent random
variables associated with a $\mathcal{G}_{\frac{9}{5}}\left( X\right) $
distribution with $\protect\beta =1$ ({\it left}), and the respective $%
\left( q=\frac{9}{5}\right) $-Fourier Transform, $\tilde{P}\left( k\right) $%
, \textit{vs.} $k$ ({\it right}). {\it Middle panels:} Same as above, in $\ln _{\frac{9}{5%
}}$-squared scale ({\it left}), and $\ln _{\frac{7}{3}}$-squared scale ({\it right}). The
straight lines indicate that $P\left( Y_{N}\right) $ and $%
\tilde{P}\left( k\right) $ are $q$-Gaussians with $q=\frac{9}{5}$ and $q=\frac{7}{3}$ respectively.
Their slopes are $\protect\beta _{q_{\ast }=9/5}^{-1}\left( N\right) $ for
left panel curves and $\protect\beta _{q_{\ast }=9/5}^{\prime }\left(
N\right) $ for right panel curves. {\it Lower panels:} $\protect\beta %
_{q_{\ast }=9/5}^{-1}\left( N\right) $ \textit{vs.} $N^{5}$, which is a
straight line with slope $1$ ({\it left}); $\protect\beta _{q_{\ast
}=9/5}^{\prime }\left( N\right) $ \textit{vs.} $N$, which is also a straight
line, but with slope $\left. \frac{3-q_{\ast }}{8\,C_{q_{\ast }}^{2\left(
q_{\ast }-1\right) }}\right\vert _{q_{\ast }=9/5}=0.030995\ldots $ ({\it right}). }
\label{clt-95}
\end{figure}

Another illustration is presented in Fig.~\ref{clt-95} for the case of the
sum of $\left( q=\frac{7}{3}\right) $-independent random variables
associated with a $\mathcal{G}_{\frac{9}{5}}\left( X\right) $ distribution
with $\beta =1$. As we have verified when the random variables have the same
probability density function but are $\left( q=1\right) $%
-independent instead, the convolution leads to a $\alpha $-stable
distribution, $L_{\alpha }\left( Y_{N\rightarrow \infty }\right) $, with $%
\alpha =\frac{3}{2}$. In the case of $\left( q=\frac{7}{3}\right) $%
-independence, the limiting (stable) distribution is the $\frac{9}{5}$-Gaussian in
accordance with the $q$-generalised central limit theorem. Consistently,
 $\beta _{q_{\ast }}^{-1}\left( N\right) =N^{5}$ and $\beta _{q_{\ast
}}^{\prime }\left( N\right) =N\,\beta _{q_{\ast }}^{\prime }\left( 1\right) $%
.

\subsection{The $\left( q,\protect\alpha\right) $-stable
distribution}

Within nonextensive statistical mechanics, the $\alpha $%
-stable L\'{e}vy distribution has also been generalised \cite{q-clt-levy}. A distribution $
f\left( x\right) $ has been considered  whose asymptotic form, $\left\vert X\right\vert
\rightarrow \infty $, corresponds to $f\left( X\right) \sim C\,\left\vert
X\right\vert ^{-\frac{1+\alpha }{1+\alpha \left( q-1\right) }}$ (for $
q=1$, $f\left( x\right) $ recovers the already mentioned $\alpha $-stable L\'{e}vy
distribution). Within this new class, we can define, 
\textit{e.g.}, a $q$-Cauchy distribution when $\alpha =1$, which leads to
the classical Cauchy distribution for $q=1$. Just as the usual L\'{e}vy
distribution, this $q$\emph{-generalisation of the }$\alpha $\emph{-stable
distribution, }$\mathcal{L}_{q,\alpha }\left( X\right) $\emph{, is defined
by its }$q$\emph{-Fourier Transform}%
\begin{equation}
\mathcal{F}_{q}\left[ \mathcal{L}_{q,\alpha }\left( X\right) \right] \left(
k\right) =C^{\prime }\exp _{q_{1}}\left[ -\beta ^{\prime }\left\vert
k\right\vert ^{\alpha }\right] ,  \label{q-alfa-levy}
\end{equation}%
where, as stated previously, $q_{1}=\frac{1+q}{3-q}=v\left( q\right) $.
Distribution $\mathcal{L}_{q,\alpha }\left( X\right) $ presents an infinite $%
\left( 2q-1\right) $-variance, when $1\le q<2$, $0<\alpha <2$, and $\alpha <%
\frac{1}{1-q}$. From this point on we denote, $\mathcal{G}_{q,\alpha }\left(
z\right) \equiv A\exp _{q}\left[ -c\left\vert z\right\vert ^{\alpha }\right] 
$. Parameter $A$ is related to the normalisation of the distribution and $c$ basically controls its width.

Along the lines of the previous generalisation, it has been shown that the
sum of $q_{+}$-independent random variables, all having the same
distribution $f\left( x\right) $, leads to Eq.~(\ref{q-alfa-levy}). Since $%
f(x)$ is \textit{stable}, \textit{i.e.}, after an appropriate scaling, the
sum of $X$ variables has the same form for the probability distribution. In other words,  
\emph{$f\left( x\right) $ is an attractor in the probability space, being a $%
\mathcal{L}$}$_{q,\alpha }\left( X\right) $. In addition, distribution $%
f\left( X\right) $ is asymptotically equivalent to $\mathcal{G}%
_{q_{L}}\left( X\right) \equiv \mathcal{G}_{q_{L},2}\left( X\right) $, where 
$q_{L}=\left( 2\,q\,\alpha -\alpha +3\right) /\left( \alpha +1\right) $. In
this way, we can say that the following mapping has been asymptotically
introduced, $\mathcal{G}_{_{q_{L},2}}\underset{\mathcal{F}_{q}}{\rightarrow }%
\mathcal{G}_{q,\alpha }\,.$

In the second paper by the same authors \cite{q-clt-levy}, an extension for
all $1 \le q<2$,$\ 0<\alpha < 2$ has been introduced. This extension is based
on the asymptotic equivalence between $f\left( X\right) $ and the $\left(
q,\alpha \right) $-distribution, $\mathcal{G}_{q,\alpha }\left( X\right)
\sim \left\vert X\right\vert ^{-\frac{\alpha }{q-1}}$ ($\left\vert
X\right\vert \rightarrow \infty $) \footnote{%
It should be emphasised that $\mathcal{G}_{q,\alpha }\left( X\right) $ 
\textit{is not} a ($q,\alpha $)-stable  distribution since it does
not respect Eq.~(\ref{q-alfa-levy}).} , and the fact that, for $0<\alpha
\leq 2$, and arbitrary $q_{1}$, there exists an index $q_{2}$, together with
a one-to-one mapping, such that $\mathcal{G}_{_{q_{1}},\alpha }\left(
X\right) \underset{\mathcal{M}_{q_{1},q_{2}}}{\rightarrow }\mathcal{G}%
_{q_{2},2}\left( X\right)$. Using this mapping, it has been obtained the
extension of Eq.~(\ref{q-fourier-transf}) for $\mathcal{G}_{q,\alpha }\left(
X\right) $, which is written in the following way 
\begin{equation}
\mathcal{G}_{q,\alpha }\underset{\mathcal{F}_{q_{\alpha }^{\ast }}}{\overset{%
\left( a\right) }{\rightarrow }}\mathcal{G}_{q^{\prime },\alpha },\qquad
1<q<2,\ 0<\alpha \leq 2 \,,  \label{fourier-q-alfa-distribution}
\end{equation}%
where 
\begin{equation}
\frac{\alpha }{1-q^{\prime }}=1+\frac{\alpha }{1-q}\,\,\,\mathrm{\ and\ }%
\,\,\,q_{\alpha }^{\ast }=\frac{\alpha -2\left( 1-q\right) }{\alpha } \,,
\label{q-attractor}
\end{equation}%
$\left( a\right) $ standing for asymptotic behaviour. Moreover, it has been
shown that this inverse $q$-Fourier transform exists, \textit{i.e.},

\begin{equation}
\mathcal{G}_{q^{\prime },\alpha }\underset{\mathcal{F}_{q_{\alpha }^{\ast
}}^{-1}}{\overset{\left( a\right) }{\rightarrow }}\mathcal{G}_{q,\alpha }.
\label{inverse-fourier-q-alfa}
\end{equation}%
The consecutive application of $n$ $q$-Fourier transforms, $\mathcal{F}%
_{q_{\alpha }^{\ast }}$, each one following Eq.~(\ref{q-attractor}), leads
to,
\begin{equation}
\frac{\alpha }{1-q_{n}}=\frac{\alpha }{1-q}+n,\qquad n=0,\pm 1,\pm 2,\ldots   \label{qn-sucessao-levy}
\end{equation}
Hence, following the previous scheme, it has also been proved \cite
{q-clt-levy} that, if we consider the sum $Y=\frac{X_{1}+X_{2}+\ldots +X_{N}}{%
D_{N}\left( q\right) }$ of symmetric variables $X_{i}$ mutually $q_{1}$-independent, 
and having a probability density function $f\left( X\right)
\sim \left\vert X\right\vert ^{-\frac{\alpha +1}{1+\alpha \left( q-1\right) }%
}$ ($\left\vert X\right\vert \rightarrow \infty $, $D_{N}\left( q\right)
\propto N^{\frac{1}{\alpha \left( 2-q\right) }}$), then, when $N\rightarrow
\infty $, $\mathcal{F}_{q}\left[ \mathcal{P}\left( Y\right) \right] \left(
k\right) =\exp _{q_{1}}\left[ -\left\vert k\right\vert ^{\alpha }\right] $,
a $\left( q_{1},\alpha \right) $-stable distribution. Bearing in mind Eq. (\ref%
{q-attractor}) and Eq.~(\ref{inverse-fourier-q-alfa}), we have that $Y$\emph{%
\ is convergent to a }$\mathcal{G}_{\left( \tilde{q},\alpha \right) }(X)$ 
\emph{distribution with}
\begin{equation}
\tilde{q}=\frac{2\left( 1-q\right) -\alpha \left( 1+q\right) }{2\left(
1-q\right) -\alpha \left( 3-q\right) }.
\end{equation}

Regarding scaling, $D_{N}(q)$, assumed for variable $Y\equiv \frac{X_{1}+X_{2}+\ldots +X_{N}}
{D_{N}\left( q \right) }$, we can define the scaling rate exponent for $\mathcal{G}_{q,\alpha }\left( X\right) $
(asymptotically equal to $f(X)$) as, $\delta =\left[ \alpha \left(
2-q\right) \right] ^{-1}$.

Summarising, we have two different approaches to the attractor of the sum of
random variables following a $\left( q,\alpha\right) $-stable distribution.
These two approaches can be understood as the existence of a crossover
between two regimes in the attractor for $\left( q,\alpha\right) $-stable
distributions. The first regime corresponds to the intermediate region of the
attractor in which it is asymptotically equal to a $\mathcal{G}_{\tilde
{q}%
,\alpha}\left( X\right) $, whose tail exponent is $\left[ 2\left( 1-q\right)
-\alpha\left( 3-q\right) \right] /\left[ 2\left( 1-q\right) \right] $. When $%
\alpha=2$, this exponent coincides with the exponent of the attractor for the
convolution of $q$-Gaussians, $2/(q-1)$. The second regime, which tends to
the $\alpha$-stable L\'evy distribution when $q=1$, represents the behaviour for
large $\left\vert X\right\vert $ where the attractor, $\mathcal{L}%
_{q,\alpha}\left( X\right) $, is asymptotically equivalent to a $\mathcal{G}%
_{q_{L},2}\left( X\right) $ distribution which has a tail exponent equal to $%
\left( \alpha+1\right) /\left( 1+\alpha\,q-\alpha \right) $. When $q=1$ this
slope coincides with the exponent for a $\alpha $-stable L\'{e}vy
distribution, $\alpha+1$. Yet in this regime, and $q=1$, we verify $q_{L}=%
\frac{\alpha+3}{\alpha+1}$, which coincides with the relation obtained in
reference \cite{ct-andre-levy}. For $\alpha\rightarrow0$, we have $%
q_{L}\rightarrow3$, upper limit for normalisation of $\mathcal{G}_{q}\left(
X\right) $, and when $\alpha\rightarrow2$, $q_{L}\rightarrow\frac{5}{3}$, 
\textit{i.e.}, the upper limit for finiteness of the second-order moment, $%
\sigma_{1}^{2}$, of $\mathcal{G}_{q}\left( X\right) $.

\begin{figure}[tbh]
\includegraphics[width=0.5\columnwidth,angle=0]{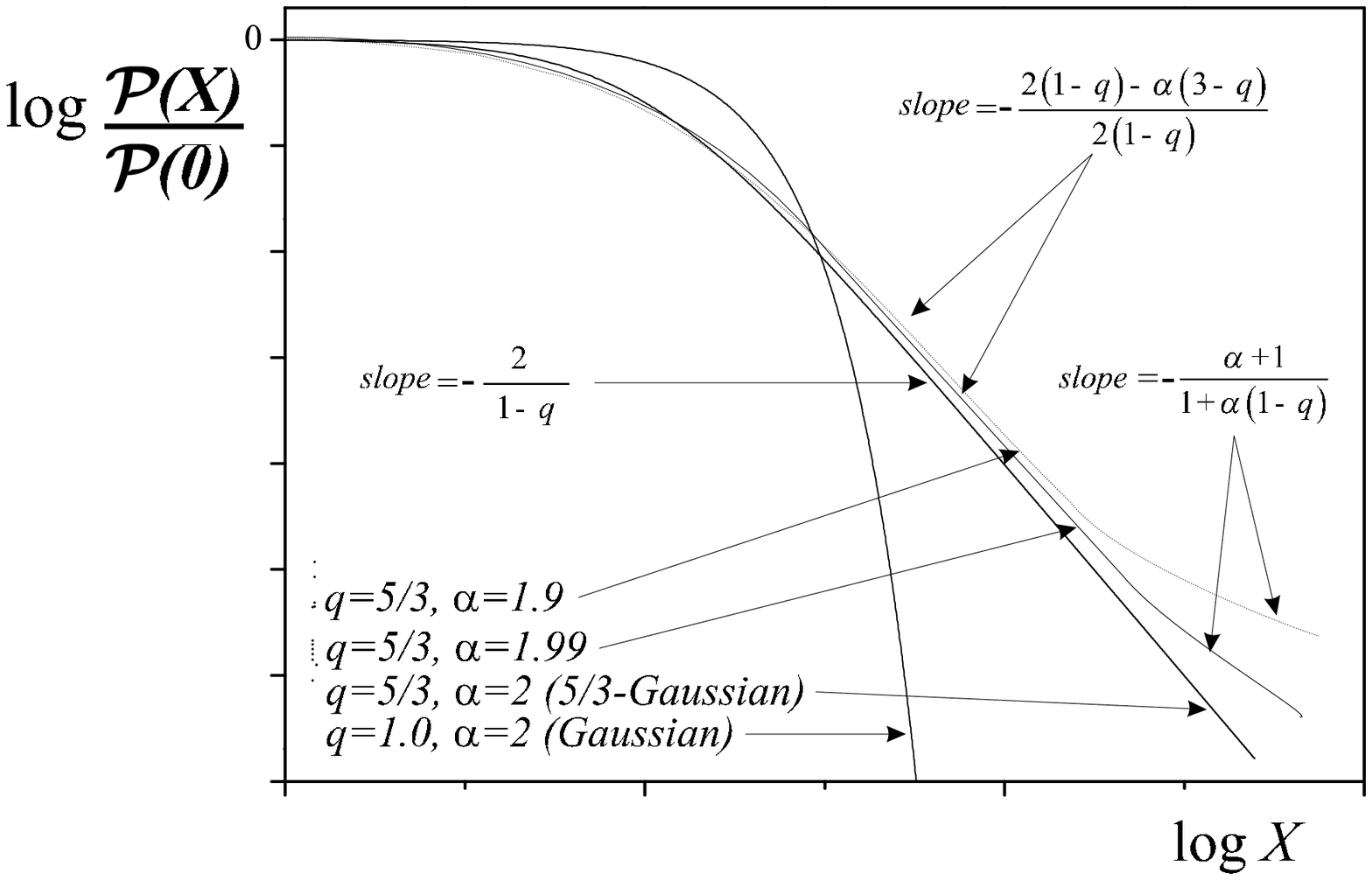} 
\caption{ Outline of $\left( q,\protect\alpha \right) $-stable distributions
for the case in which the correlation is given by $q_{1}=2$. As $\protect%
\alpha $ of L\'{e}vy distributions approaches $2$, the distributions $\left(
q,\protect\alpha \right) $-stable becomes more and more similar to a $G_{%
\frac{5}{3}}\left( X\right) $ with an exponent $\left[ 2\left( 1-q\right) -%
\protect\alpha \left( 3-q\right) \right] /\left[ 2\left( 1-q\right) \right] $%
. However, since $\protect\alpha \neq 2$, for some critical value $X^{\ast }$%
, the distribution changes to another regime with a tail exponent  $\left( 
\protect\alpha +1\right) /\left( 1+\protect\alpha \,q-\protect\alpha \right) 
$. }
\label{sketch}
\end{figure}

The former analysis is evocative of Fig.~5 of Ref. \cite{ct-smdq-1}. In
other words, as $\alpha \rightarrow 2$, $\mathcal{L}_{q,\alpha }\left(
X\right) $ approaches the $q$-Gaussian, $\mathcal{G}_{q}\left( X\right) $
and, at some critical value $X_{c}$, it changes its behaviour to a power-law
decay with exponent $\left( \alpha +1\right) /\left( 1+\alpha \,q-\alpha
\right) $. With this picture in mind, we sketch in Fig.~\ref
{sketch} the attractor for the case $q_{1}=2=v\left( q=\frac{%
5}{3}\right) $ and values of $\alpha $ approaching $\alpha =2$. We can
verify that, for all $1<q<2$, the inequalities $\frac{2}{q-1}\geq 
\frac{2\left( 1-q\right) -\alpha \left( 3-q\right) }{2\left( 1-q\right) }>%
\frac{\alpha +1}{1+\alpha \left( q-1\right) }$ hold. Our sketch might be
corroborated in the near future, as soon as the form of
the inverse $q$-Fourier Transform becomes analytically available. A summary of the whole situation is presented in
Table~\ref{tabela}.

\begin{table}[ptb]
\caption{\textit{R\'{e}sum\'{e}} of the main results presented in the
article: Central limit theorems which present a $N^{1/\left[ \protect\alpha %
\left( 2-q\right) \right] }$-scaled attractor $\mathbb{F}\left( X\right) $
for the sum of $N\rightarrow \infty $ $q$-independent identical random
variables with symmetric distribution $f\left( X\right) $;
$q_{1}\equiv \frac{q+1}{3-q}$; we focus on $q \ge 1$. The term \textit{intermediate}
must be read as an infinity, however not so large as the infinity associated
with the \textit{distant} regime. For $q \neq 1$ and $\protect\alpha =2$,
when the random variables are associated with a $q$-Gaussian, we verify the
scaling relation, $\protect\beta ^{\prime }(N)=N\protect\beta ^{\prime }$,
where $\protect\beta ^{\prime }$ is the (inverse) width for the $q$-Fourier
Transform.}
\label{tabela}%
\begin{tabular}{l|c|c|}
\cline{2-3}
& $%
\begin{array}{c}
\text{ \ } \\ 
q=1\text{ \ [independent]} \\ 
\text{ \ }%
\end{array}%
$ & $%
\begin{array}{c}
\text{ \ } \\ 
q \neq 1\text{ [globally correlated]} \\ 
\text{ \ }%
\end{array}%
$ \\ \hline
\multicolumn{1}{|l|}{$%
\begin{array}{c}
\sigma _{2q-1}<\infty \\ 
\left( \alpha =2\right)%
\end{array}%
$} & $%
\begin{array}{c}
\mathbb{F}\left( X\right) =\mathcal{G}\left( X\right) \\ 
\text{\textrm{with\ same\ }}\sigma _{1}\text{\ \textrm{of\ }\/}f\left(
X\right) \\ 
\\ 
\text{Classical CLT}%
\end{array}%
$ & \multicolumn{1}{|l|}{$%
\begin{array}{c}
\text{ \ } \\ 
\mathbb{F}\left( X\right) =\mathcal{G}_{q}\left( X\right) =\mathcal{G}_{%
\frac{3\,q_{1}-1}{1+q_{1}}}\left( X\right) \qquad \mathrm{stable\
distribution} \\ 
\left[ \text{\textrm{with\ same\ }}\sigma _{2\,q-1}\text{\ \textrm{of\ }\/}%
f\left( X\right) \right] \\ 
\\ 
\mathcal{G}_{q}\left( X\right) \sim \left\{ 
\begin{array}{l}
\mathcal{G}\left( X\right) \qquad \qquad \qquad \,\,\,\mathrm{if}\ \
\left\vert X\right\vert \ll X_{c}\left( q,2\right) ; \\ 
\\ 
C_{q,2}/\left\vert X\right\vert ^{2/\left( q-1\right) }\qquad \mathrm{if}\ \
\left\vert X\right\vert \gg X_{c}\left( q,2\right)%
\end{array}%
\right. \\ 
\text{for }q>1\text{, with }\lim_{q\rightarrow 1}X_{c}\left( q,2\right)
=\infty \\ 
\text{ \ }%
\end{array}%
$} \\ \hline
\multicolumn{1}{|l|}{$%
\begin{array}{c}
\sigma _{2q-1}\rightarrow \infty \\ 
\left( 0<\alpha <2\right)%
\end{array}%
$} & \multicolumn{1}{|l|}{$%
\begin{array}{c}
\text{ \ } \\ 
\mathbb{F}\left( X\right) =L_{\alpha }\left( X\right) \qquad \mathrm{stable\
distribution} \\ 
\left[ \text{\textrm{with\ same\ }}\left\vert X\right\vert \rightarrow
\infty \ \mathrm{behaviour}\text{ \textrm{of\ }\/}f\left( X\right) \right]
\\ 
\\ 
L_{\alpha }\left( X\right) \sim \left\{ 
\begin{array}{l}
\mathcal{G}\left( X\right) \qquad \mathrm{if}\ \ \left\vert X\right\vert \ll
X_{c}\left( 1,\alpha \right) ; \\ 
\\ 
C_{1,\alpha }/\left\vert X\right\vert ^{\alpha +1}\qquad \mathrm{if}\ \
\left\vert X\right\vert \gg X_{c}\left( q,2\right)%
\end{array}%
\right. \\ 
\text{with }\lim_{\alpha \rightarrow 2}X_{c}\left( 1,\alpha \right) =\infty
\\ 
\text{ \ } \\ 
\mathrm{L\acute{e}vy-Gnedenko\ CLT}%
\end{array}%
$} & \multicolumn{1}{|l|}{$%
\begin{array}{c}
\text{ \ } \\ 
\mathbb{F}\left( X\right) =\mathcal{L}_{q,\alpha }\left( X\right) \qquad 
\mathrm{stable\ distribution} \\ 
\text{ \ }\left[ \text{\textrm{with\ same\ }}\left\vert X\right\vert
\rightarrow \infty \ \mathrm{behaviour}\text{ \textrm{of\ }\/}f\left(
X\right) \right] \\ 
\\ 
\mathcal{L}_{q,\alpha }\left( X\right) \sim \left\{ 
\begin{array}{c}
\mathcal{G}_{\frac{2\left( 1-q\right) -\alpha \left( 1+q\right) }{2\left(
1-q\right) -\alpha \left( 3-q\right) },\alpha }\left( X\right) \sim
C_{q,\alpha }^{\ast }/\left\vert X\right\vert ^{\frac{2\left( 1-q\right)
-\alpha \left( 3-q\right) }{2\left( 1-q\right) }} \\ 
\text{intermediate regime } \\ 
\qquad \qquad \left[ X_{c}^{\left( 1\right) }\left( q,\alpha \right) \ll
\left\vert X\right\vert \ll X_{c}^{\left( 2\right) }\left( q,\alpha \right) %
\right] ; \\ 
\multicolumn{1}{r}{} \\ 
\mathcal{G}_{\frac{2\,\alpha \,q-\alpha +3}{\alpha +1},2}\left( X\right)
\sim C_{q,\alpha }^{L}/\left\vert X\right\vert ^{\frac{\alpha +1}{1+\alpha
\left( q-1\right) }} \\ 
\text{distant regime} \\ 
\left[ \left\vert X\right\vert \gg X_{c}^{\left( 2\right) }\left( q,\alpha
\right) \right]%
\end{array}%
\right. \\ 
\text{ \ }%
\end{array}%
$} \\ \hline
\end{tabular}%
\end{table}

\section{Final Remarks}

\label{conclusion}

In this article we have reviewed the generalised central limit theorems
presented in \cite{q-clt-gauss,q-clt-levy}. These
theorems are based on nonextensive statistical
mechanics and they address the sum of $q_{1}$-correlated random variables (with $q_{1}=\frac{1+q}{3-q}$)
whose attractor scales as $N^{-1/\left[ \alpha \left( 2-q\right) \right] }$. Introducing an appropriate nonlinear
generalisation for the Fourier Transform, it has been possible to verify the
existence of a new attractive subspace in probability space, namely $%
\mathcal{G}_{q,\alpha }\left( X\right) $, which contains the Gaussian, $%
\mathcal{G}\left( X\right) \equiv \mathcal{G}_{1,2}\left( X\right) $, and
(in an asymptotic way) $\alpha $-stable L\'{e}vy distributions, $\mathcal{L}%
_{1,\alpha }\left( X\right) \equiv \mathcal{F}_{1}^{-1}\left[ \mathcal{G}%
_{1,\alpha }\left( k\right) \right] \left( X\right) $. These special
attractors are both related to the sum of independent variables. Within $%
\mathcal{G}_{q,\alpha }\left( X\right) $, considering $q_{+}$-correlated
variables (Eq.~(\ref{q-convolution})) sharing the same distribution, and
presenting a finite $\left( 2q-1\right) $-variance, $\sigma _{2\,q-1}^{2}$,
it is possible to observe the existence of a line of attraction, $\mathcal{G}%
_{q}\left( X\right) $, ($\alpha =2$) in $q-\alpha $ space. We have also
dealt with a $q_{+}$-generalisation of the $\alpha $-stable L\'{e}vy
distribution, $\mathcal{L}_{q,\alpha }\left( X\right) \equiv \mathcal{F}%
_{q}^{-1}\left[ \mathcal{G}_{q,\alpha }\left( k\right) \right] \left(
X\right) $. Since $\mathcal{L}_{q,\alpha }\left( X\right) $ is stable, the convolution of such distributions, assuming they have an infinite $\left(
2q-1\right) $-variance, converges towards a $\mathcal{L}_{q,\alpha }\left(
X\right) $ distribution, which for large $\left\vert X\right\vert $, is
equivalent to a $\mathcal{G}_{q_{L},2}\left( X\right) $ distribution with $%
q_{L}=\frac{2\,q\,\alpha -\alpha +3}{\alpha +1}$. Removing the restriction
of infinite $\left( 2q-1\right) $-variance, the convolution of $\mathcal{L}%
_{q,\alpha }\left( X\right) $ distributions, asymptotically equivalent to $%
\mathcal{G}_{q^{\ddag },\alpha }\left( X\right) \sim \left\vert X\right\vert
^{-\frac{\alpha }{q^{\ddag }-1}}$ ($\left\vert X\right\vert \rightarrow
\infty $ and $q^{\ddag }=\frac{1+2\alpha +\alpha ^{2}\left( q-1\right) }{%
1+\alpha }$), leads to a $\mathcal{G}_{\tilde{q},\alpha }\left( X\right) $
distribution for which $\tilde{q}=\frac{2\left( 1-q\right) -\alpha \left(
1+q\right) }{2\left( 1-q\right) -\alpha \left( 3-q\right) }$. These two
asymptotic laws for $\mathcal{L}_{q,\alpha }\left( X\right) $ correspond to the
existence of a crossover that we have depicted in Fig.~\ref{sketch}.

In both cases referred above, {\it i.e.}, addition of random variables with finite and incommensurable standard deviation, it is possible to define
a triplet of parameters which characterises the attractor, the correlation,
and the scaling rate, $\left\{ P_{att},P_{cor,}P_{scl}\right\} $~\cite{q-clt-levy} reminiscent
of the $q$-triplet $\left\{ q_{sen},q_{rel},q_{stat}\right\} $ conjectured
in Ref.~\cite{tsallis-vila}. In that article, it was proposed that the values of 
$\left\{ q_{sen},q_{rel},q_{stat}\right\} $ for say systems like long-range
Hamiltonian systems characterised by the interaction-decay exponent $\alpha $ and the
dimension $d$ would respect inequalities such as
$q_{rel},q_{sta}>1$ and $q_{sen}<1$. Considering the convolution of $q$-independent random variables associated 
with $\mathcal{G}_{q_{-1}}\left( X\right) $, it has been shown that the triplet of parameters $\left\{ P_{att},P_{cor,}P_{scl}\right\} $ 
corresponds in fact to $\left\{ \frac{3q-1}{1+q},q,\frac{1+q}{3-q}\right\} $, 
or simply $\left\{q_{k-1},q_{k},q_{k+1}\right\} $ following mapping relations. 
In this way, we can replace $P$, that stands for {\it parameter}, by $q$ used
to represent entropic indices. Hence, the triplet of parameters in actualy leading to the $q$-triplet $\left\{ q_{att},q_{corr},q_{scal}\right\} $.
Establishing a bridge between the triplet porposed in Ref.~\cite{tsallis-vila} and the triplet we have discussed, we argue that $q_{k-1}$, 
the entropic index for the attractor, should equal $q_{stat}$. Taking into account Eq.~(\ref{qn-sucessao}), we can
write 
\begin{equation}
q_{k-1}=2-\frac{1}{q_{k+1}}.  \label{paper-relation}
\end{equation}%
It is noteworhy that the $q$-triplet conjecture~\cite{tsallis-vila} was first observed by NASA
using data from Voyager 1 related to the solar wind at the distant
heliosphere~\cite{nasa}, and also in a paradigmatic complex system
such as a financial market~\cite{canberra}. For the solar wind observations, it has been
inferred the relations $q_{stat}+1/q_{rel}=2,$ and $q_{rel}+1/q_{sen}=2$,
which are consistent, within experimental error, with the results obtained
from the NASA data set. Again, for reasons presented above, $q_{stat}=q_{k-1}$.
Using Eq. (\ref{paper-relation}) we have $q_{rel}=q_{k+1}$, and $%
q_{sen}=q_{k+3}$. These two cases of correspondence between $\left\{
q_{stat},q_{rel},q_{sen}\right\} $ and $\left\{ q_{k^{\ast
}},q_{k^{+}},q_{k^{\prime }}\right\} $ represent strong candidates for
the description of the $q$-triplets for conservative and dissipative
systems. 

In addition, let us also mention
two transformations that appear quite often in problems discussed
within nonextensive concepts, namely $q_{a}(q)=2-q$ and $q_{m}(q)=1/q$. In fact, these transformations, 
{\it additive and multiplicative dualities} respectively, have shown to be at the basis of
the relations between entropic indices. Explicitly, if we apply both
transformations $n$ times in a row, we obtain $\left[ q_{a}q_{m}\right] ^{n}(q)=
\frac{q+n\left( 1-q\right) }{1+n\left( 1-q\right) }$. Looking to Eq.~(\ref%
{qn-sucessao-levy}), we notice that sequences with $\alpha =2$
and $n=0,\pm 2,\pm 4,\ldots $, or $\alpha =1$ and $n=0,\pm 1,\pm 2,\ldots $
coincide with $\left[ q_{a}q_{m}\right] ^{n}$. This fact is quite remarkable 
since it reveals a connection between the sequences that emerge from the $q$%
-generalised central limit theorems and the dualities presented here above.
The physical interpretation of these as well as other relations between
entropic indices constitutes an interesting open challenge.

Finally, let us mention the fact that the Central Limit Theorem results appear to also be
applicable to the sum of deterministic variables~\cite{kaminska}. It has been
proved that, when the maximum Lyapunov exponent is positive, the sum of
deterministic variables obtained from some dynamical process leads to a
Gaussian distribution. Recently, studies on the sum of deterministic variables
obtained from dissipative and conservative systems have been made. For the
logistic map~\cite{ubt} (dissipative system) at the edge of the chaos
(vanishing Lyapunov exponent) it has been verified that the intermediate
part and the tail of the distribution are consistent with a $q$-Gaussian
distribution. Concerning conservative systems, studies on the Hamiltonian
Mean Field~model \cite{prt} at its metastable state have shown the emergence
of non-Gaussian attracting probability density functions when the sum of
velocities is performed. These resulting distributions have been numerically
quite well approached by $q$-Gaussians over its whole range of values. Further
analysis of both conservative and dissipative systems might clarify the
emergence of a new generalisation of the central limit theorem, but for
deterministic variables.

\begin{theacknowledgments}
We are deeply thankful to S. Umarov and L.G. Moyano for fruitful
discussions. Partial financial support from Pronex, CNPq, Faperj (Brazilian
agencies) and FCT/MCES (Portuguese agency) is acknowledged as well.
\end{theacknowledgments}

\bibliographystyle{aipprocl} 


\end{document}